\documentclass[a4paper,10pt,twoside]{cpc-hepnp}

\usepackage{multicol}
\usepackage{graphicx}
\usepackage{booktabs}
\usepackage{amssymb,bm,mathrsfs,bbm,amscd}
\usepackage[tbtags]{amsmath}
\usepackage{lastpage}
\usepackage{subfigure}
\usepackage{placeins}
\usepackage{CJK}

\begin{document}

%\begin{CJK*}{GB}{song}

\fancyhead[co]{\footnotesize Sun Yong-Jie et al: Beam Test of Signal Cross-talk and Transmission for LMPRC}

\footnotetext[0]{Received 10 November 2010}

\title{Beam test of signal cross-talk and transmission for LMPRC\thanks{Supported by the National Natural Science
Foundation of China~(10775131),(10875120),and the Major State Basic Research Development Program~(2008CB817702) }}

\author{%
      Sun Yong-Jie(ËïÓ½Ü)
\quad Li Cheng$^{1)}$(Àî³Î)\email{licheng@ustc.edu.cn}%
\quad Tang Ze-Bo(ÌÆÔó²¨)
\quad Xu Lai-Lin(ÐìÀ´ÁÖ)
\quad Chen Tian-Xiang(³ÂÌìÏè)
\quad Shao Ming(ÉÛÃ÷) 
}
\maketitle

\address{%
 Department of Modern Physics,  University of Science and Technology of China~(USTC),  Hefei 230026,  China\\
}

\begin{abstract}
A new prototype of large area Multi-gap Resistive Plate Chamber~(MRPC) with long readout strips was built.
This Long-strip Multi-gap Resistive Plate Chamber~(LMRPC) is double stacked and has ten 250~$\mu$m-thick gas gaps.
Signals are read out from two ends of strip with an active area of 50~cm$\times$2.5~cm in each. The detector was tested
at FOPI in GSI, using the secondary particles of proton beams~($E = 3.5~GeV$) colliding with a Pb target.
The results show that the LMRPC prototype has a time resolution of about 60$\sim$70~ps; the detecting efficiency
is over 98\% and the ratio of cross-talk is lower than 2\%. The detector also has a good spatial resolution of
0.36~cm along the strip direction.
\end{abstract}



\begin{multicols}{2}

\section{Introduction}

Multi-gap Resistive Plate Chamber~(MRPC), as a gaseous detector working at avalanche mode, has excellent time resolution
and high detecting efficiency\cite{lab1}. Experimental study shows that the intrinsic time resolution of MRPC can reach below 50~ps
and the detecting efficiency for MIPs can be nearly 100\%. Additionally, the cost of the detector is relatively low and it's
easy to construct, thus MRPC is a preferred choice for Time-Of-Flight~(TOF) detectors with large area in modern high energy
experiments, such as the STAR~(the Solenoidal Tracker at RHIC)~TOF systems at RHIC\cite{lab2}~(the Relativistic Heavy Ion Collider) and the ALICE~(a Large Ion Collider Experiment)~TOF systems at LHC\cite{lab3}~(the Large Hadron Collider). The readout cells are
flexible to shape into pads or strips to achieve the specific requirements of the experiment. In the case of low multiplicities of charged
particles, large area MRPCs with long strips readout~(LMRPC) are superior for the following advantages:~1)~with the optimization
of chamber design and signal readout,  LMPRC can reduce the number of electronics channels and save the cost further;~2)~when
readout signals are from two ends of the strip , the average timing of two ends is more accurate, and the time jitter caused by the difference of hit positions on the strip can be eliminated;~3)~using ``left minus right''timing, this readout method can provide a measurement of the hit position along the strip.

The superiority of this technology of LMRPC has been proven by experiment and it will be widely used in future high energy physics experiments. The first LMRPC prototypes designed by USTC~HEP group for STAR~MTD~(Muon Telescope Detector) have an active area of 87.0$\times$17.0~$cm^{2}$, and they have achieved a time resolution of 60$\sim$70~ps and a spatial resolution of $\sim$1~cm along the strip direction\cite{lab4}. Recently a new prototype with much larger area of 89.0$\times$55.0~$cm^{2}$ has been proposed for MTD detector. The CBM~(the Compressed Baryonic Matter experiment) experiment at FAIR~(Facility for Antiproton and Ion Research) also plans to use 15$\sim$150~cm-long LMRPC detectors as a part of their TOF~Wall system\cite{lab5}.

However, the long readout strip might also bring about an undesired side effect, known as cross-talk. The longer the readout strips, the stronger the electromagnetic interference between adjacent strips might appear. Cross-talk can affect the LMRPC's performance, such as the time resolution and the counting capability, especially in the environment of high multiplicity which is common in the heavy ion collision experiments. Therefore, cross-talk is an important issue for LMRPC  before its future application.

For this purpose, a new LMRPC prototype with an active area of 50.0$\times$15.0~$cm^{2}$ was designed and its performance, especially the time resolution and signal cross-talk were studied in detail.

\section{The LMRPC prototypes}

\begin{figure*}[!ht]

\subfigure[ ]{
    \label{fig1:subfig:a}
    \begin{minipage}[b]{0.5\textwidth}
    \centering
    \includegraphics[width=12cm]{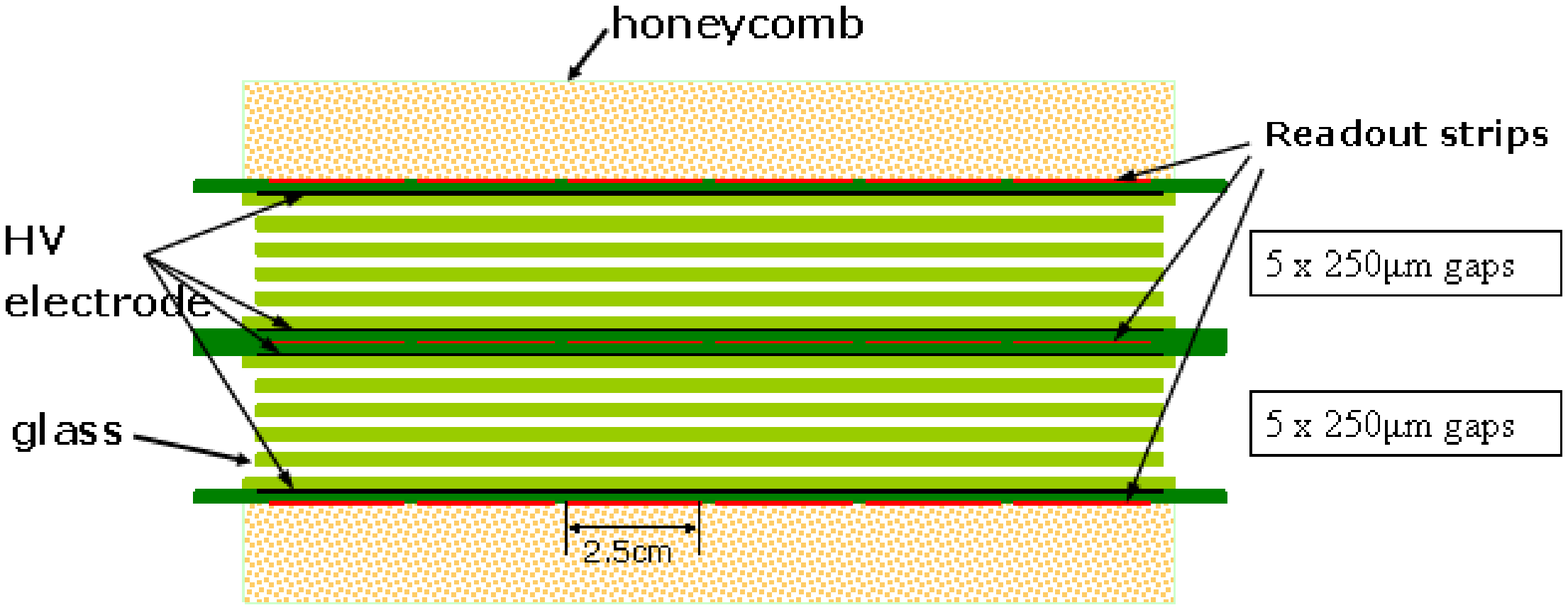}
    \end{minipage}}

\subfigure[]{
    \label{fig1:subfig:b}
    \begin{minipage}[b]{0.5\textwidth}
    \centering
    \includegraphics[width=9cm]{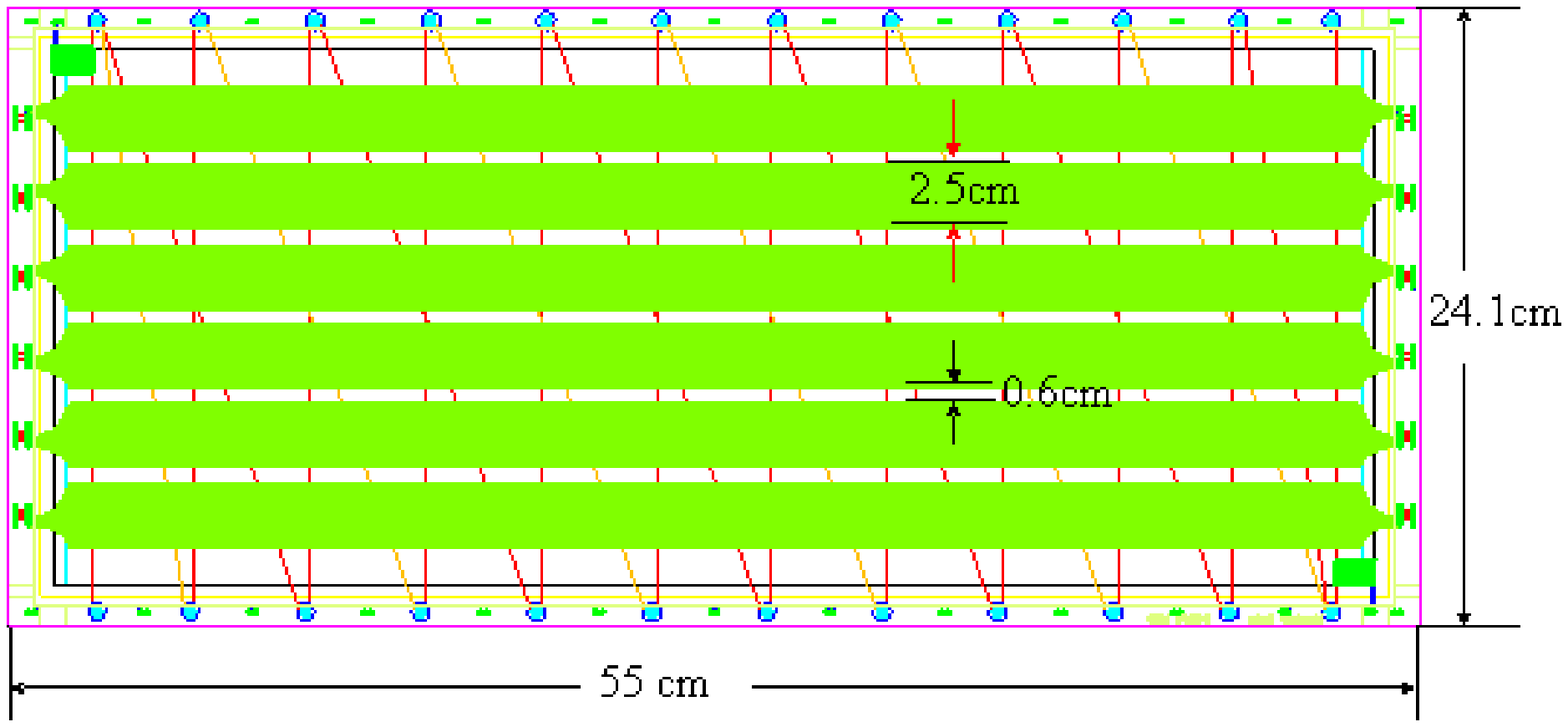}
    \end{minipage}}
\caption{(a)Schematic side-view of the LMRPC module. Ten gas gaps of 250~$\mu$m wide are arranged in two stacks. High voltage is applied on the surfaces of the outmost glass of each stack. (b)The LMRPC structure from top-vies. The readout strips located on the printed circuit board. The strips are 2.5 cm wide. The gaps between strips are 0.6 cm. (The red zigzag line represents the path of the fishing line.)}

\end{figure*}

Fig.~\ref{fig1:subfig:a} schematically shows the structure of the double-stack LMRPC module with 10 gas gaps. Each gap is 250~$\mu$m-thick and is defined using a nylon monofilament fishing line. This LMRPC has resistive plates made of 0.7~$mm$ float glass sheets with volume resistivity of$\sim$$10^{13}$$\Omega$$\cdot$~cm.The electrodes are made of Licron crystals coating
sprayed on the outermost glasses of each stack, with resistivity of$\sim$40$M\Omega$/$\Box$. The readout pads are segmented into six double-ended strips that are 2.5~cm wide and 50~cm long, as shown in Fig.\ref{fig1:subfig:b}. The gap between each strip is 6~$mm$ wide. Thus the active area of this LMRPC module is 50.0$\times$18.0~$cm^{2}$. Pins are used to transmit the negative signals from the external strip layers to the middle strip layers, thus differential signals are collected from both ends of the strips and are sent to the front-end electronics. These pins can also fix
the three boards together. The LMRPC module is enclosed in a gas-tight aluminium box. A gas mixture which contains 95\% Freon R-134a,
3\% $iso-C_{4}H_{10}$ and 2\% $SF_{6}$  flows as working gas.

\end{multicols}

\begin{multicols}{2}

\section{Beam test setup}

\begin{figure*}[!ht]
  \subfigure[]{
    \label{fig2:subfig:a} %% label for first subfigure
    \begin{minipage}[b]{0.5\textwidth}
      \centering
      \includegraphics[width=1\textwidth]{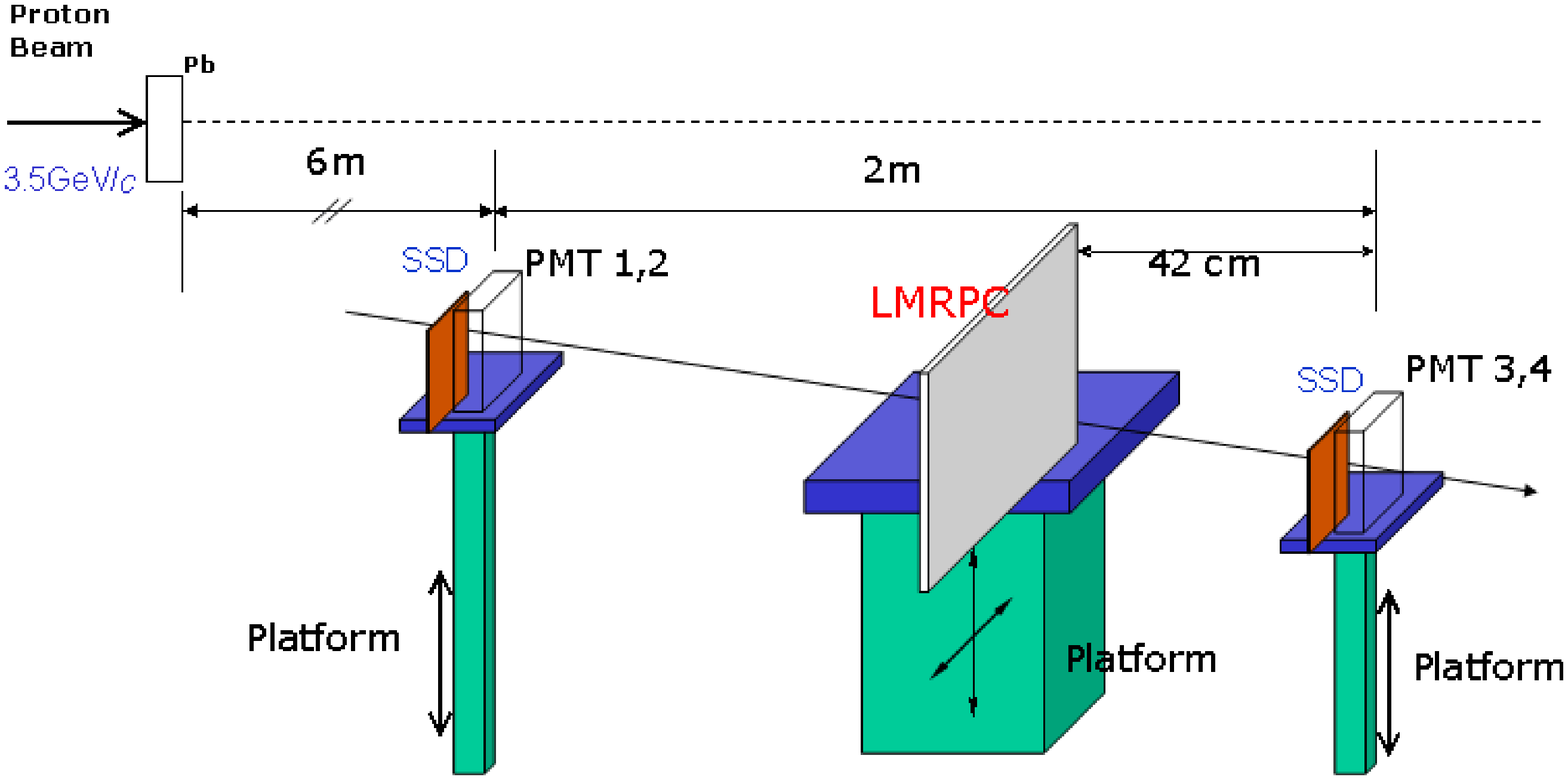}
    \end{minipage}}%
  \subfigure[]{
    \label{fig2:subfig:b} %% label for second subfigure
    \begin{minipage}[b]{0.5\textwidth}
      \centering
      \includegraphics[width=0.8\textwidth]{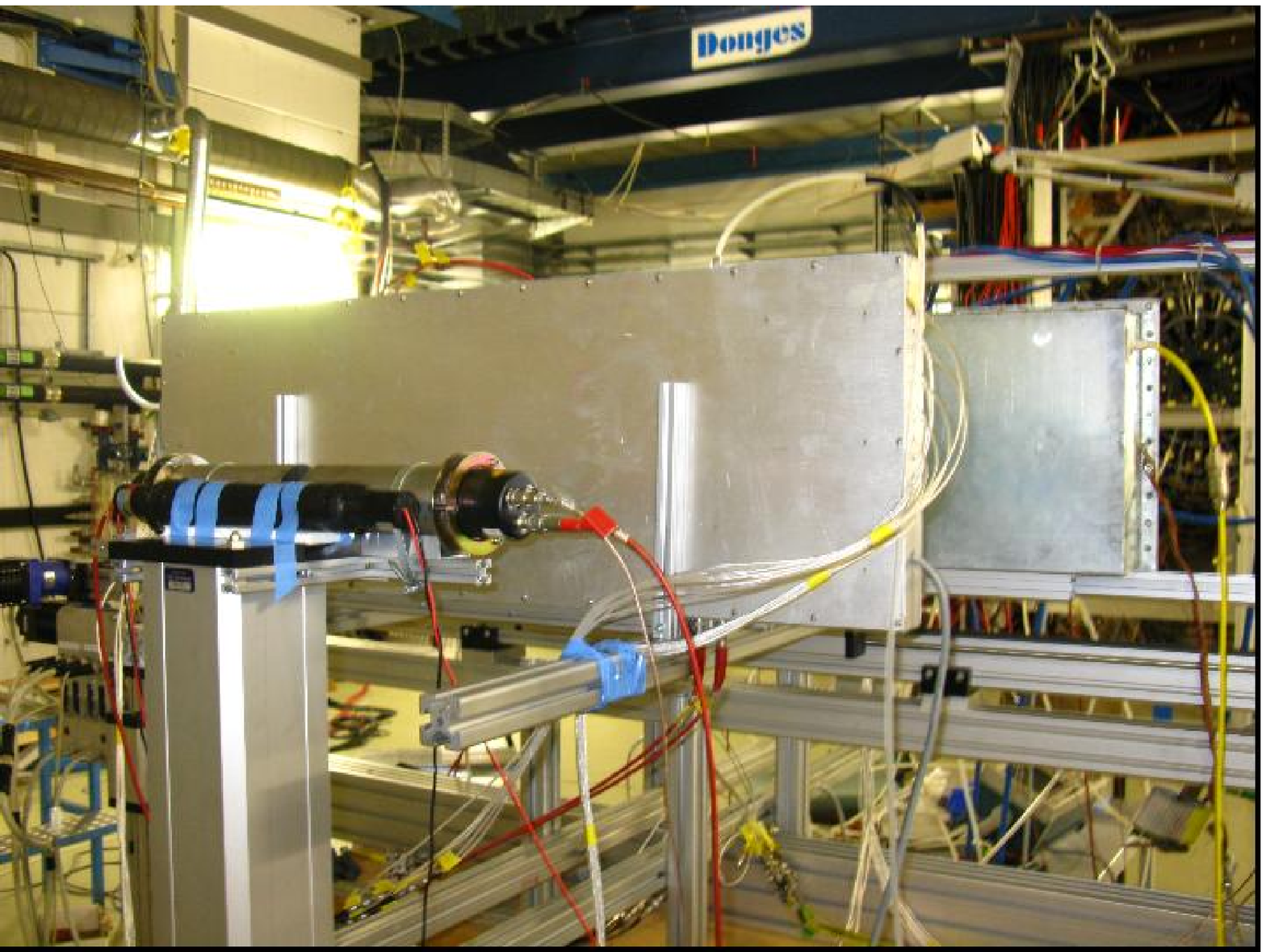}
    \end{minipage}}
  \subfigure[]{
    \label{fig2:subfig:c}
    \begin{minipage}[b]{0.5\textwidth}
      \centering
      \includegraphics[width=7cm]{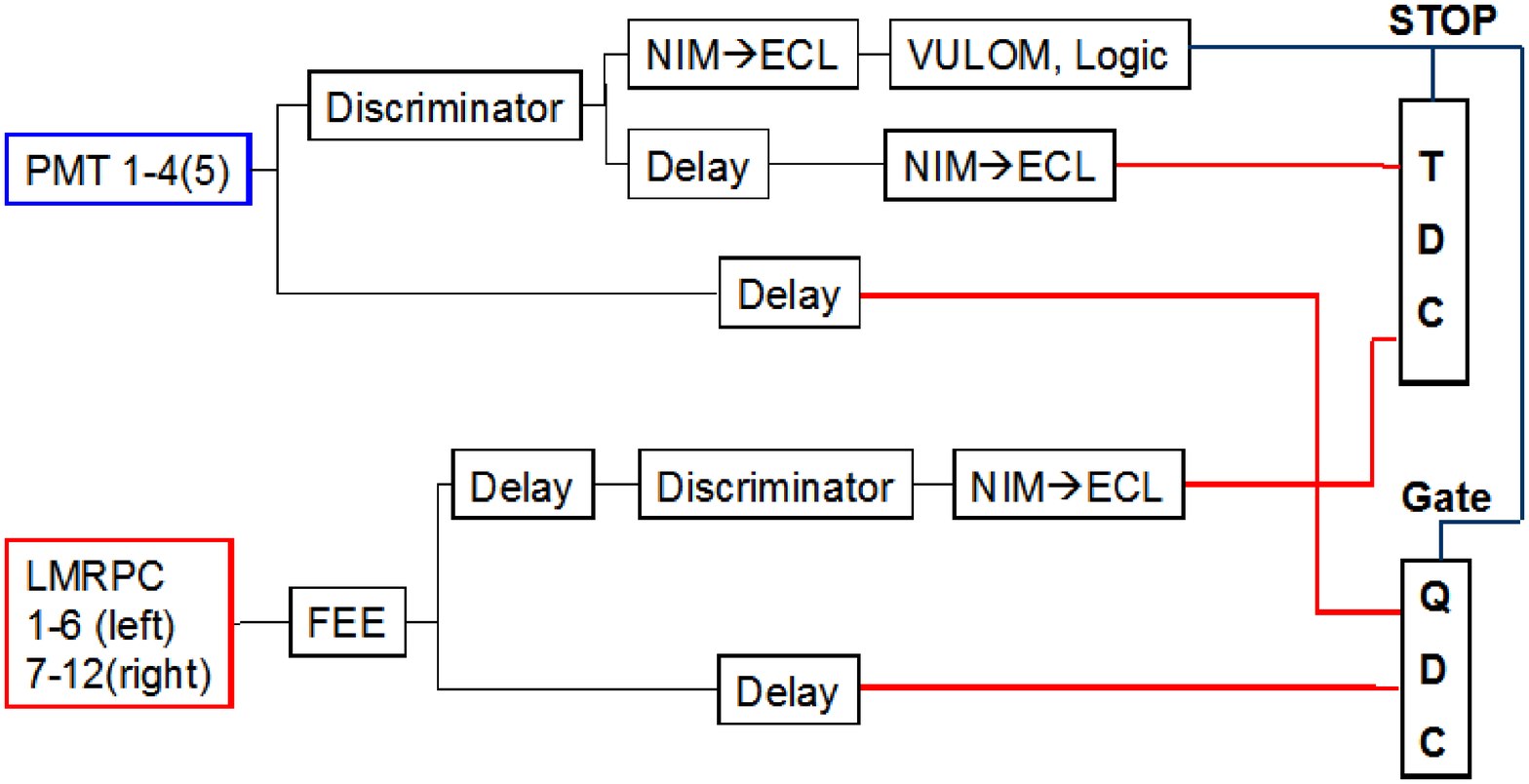}
    \end{minipage}}

\caption{(a)The schematic of the beam test experiment setup. (b)The picture of the setup, taken from the downstream of the beam. (c)The data acquiring logic.}

\end{figure*}

A beam test experiment was carried out at the FOPI~(4 $\pi$) facility at GSI~(GSI Helmholtz Centre for Heavy Ion Research GmbH) in August 2009. The test beam consistes of the secondary particles at a fixed angle from 3.5~GeV protons colliding with a lead target. The setup of the beam test system is shown in Fig.\ref{fig2:subfig:a}. Two scintillators were placed at the up- and down-stream of the LMRPC to make a telescope detecting system. The telescope system defined the trigger area of 2~cm~(in y-direction)$\times$4~cm~(in x-direction, along the strip). The strips of the LMRPC were along x direction. These devices were placed on movable platforms. Each scintillator was read out via two photomultiplier tubes~(PMT) at the two ends~(PMT1, PMT2 for the upstream scintillator and PMT3, PMT4 for the downstream scintillator). They provided the reference time $T_0$ and the coincidence of the four PMTs was used as the event trigger. This trigger was sent to the QDCs as the gate signals and to the TDCs as the common stop signals, as shown in Fig.\ref{fig2:subfig:c}.

\end{multicols}

\begin{multicols}{2}

\section{Experimental results}

The resolution of the reference time $T_0$ is shown in Fig.\ref{fig3}, where $T_0$ is defined as the average time of the four PMTs. The resolution of $T_0$ after slewing correction is 67~ps. The details of slewing correction will be discussed later.
Fig.\ref{fig4} shows the LMRPC's plateau of detecting efficiency versus high voltage~(HV). The time resolution as a
function of HV is also plotted in this figure. The detecting efficiency, which is defined as both ends of any strip
have valid signals, is above 98\% for the positive and negative HV greater than 6.9~kV.  Removing the time jitter of
the reference time, the time resolution of LMRPC is about 60$\sim$70~ps. Since the test beam consists of all the secondary particles, this time resolution also includes the contribution of the momentum dispersion.

\begin{center}
\setlength{\belowcaptionskip}{10pt}
\includegraphics[width=7cm]{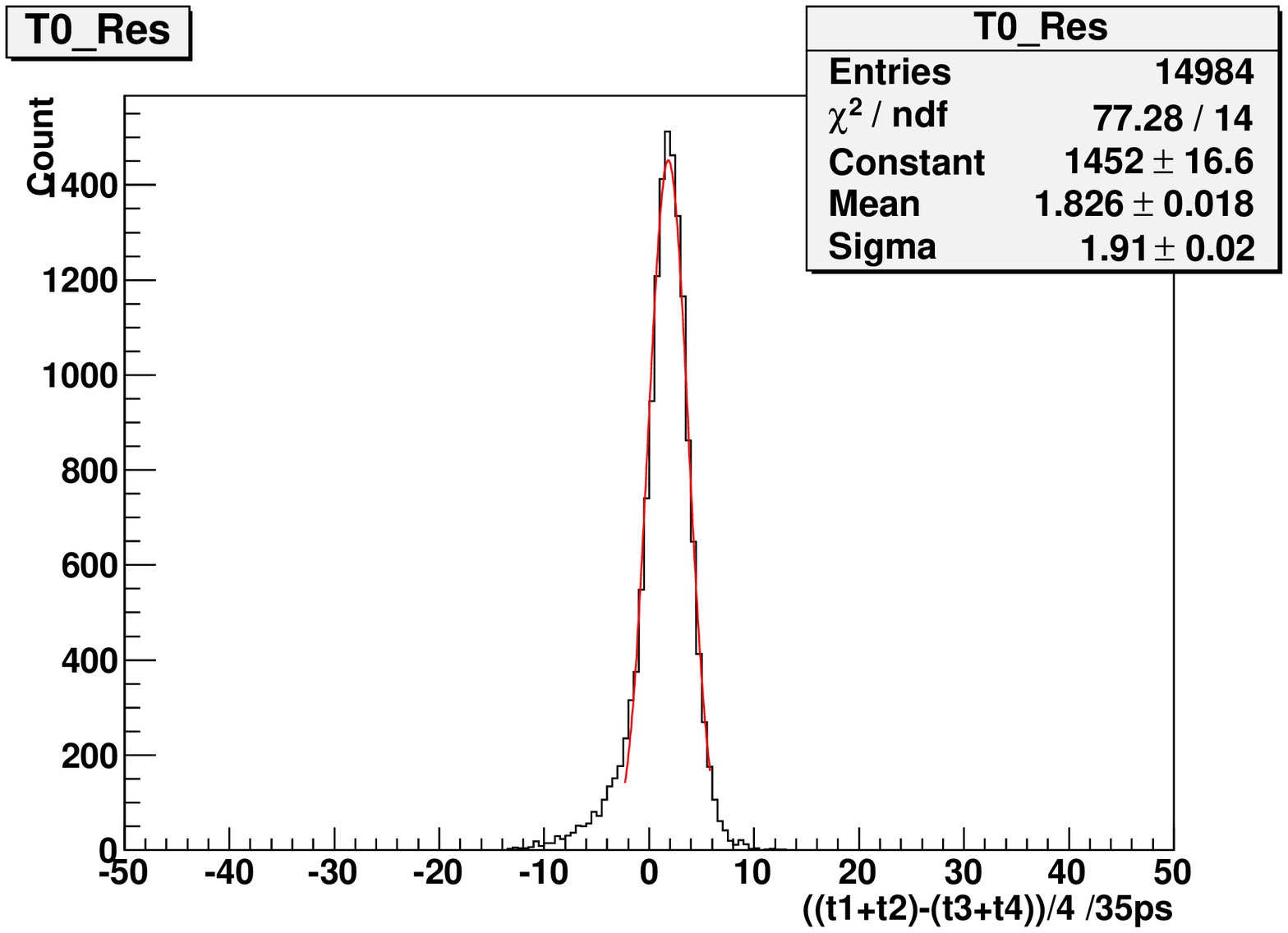}
\figcaption{\label{fig3}   The distribution of reference time $T_0$, where $T_0=((t_1+t_2)-(t_3+t_4))/4$, $t_i$ stands for the time measured by 4 PMTs and the unit is 35~ps per TDC channel.  }
\end{center}

\begin{center}
\includegraphics[width=7cm]{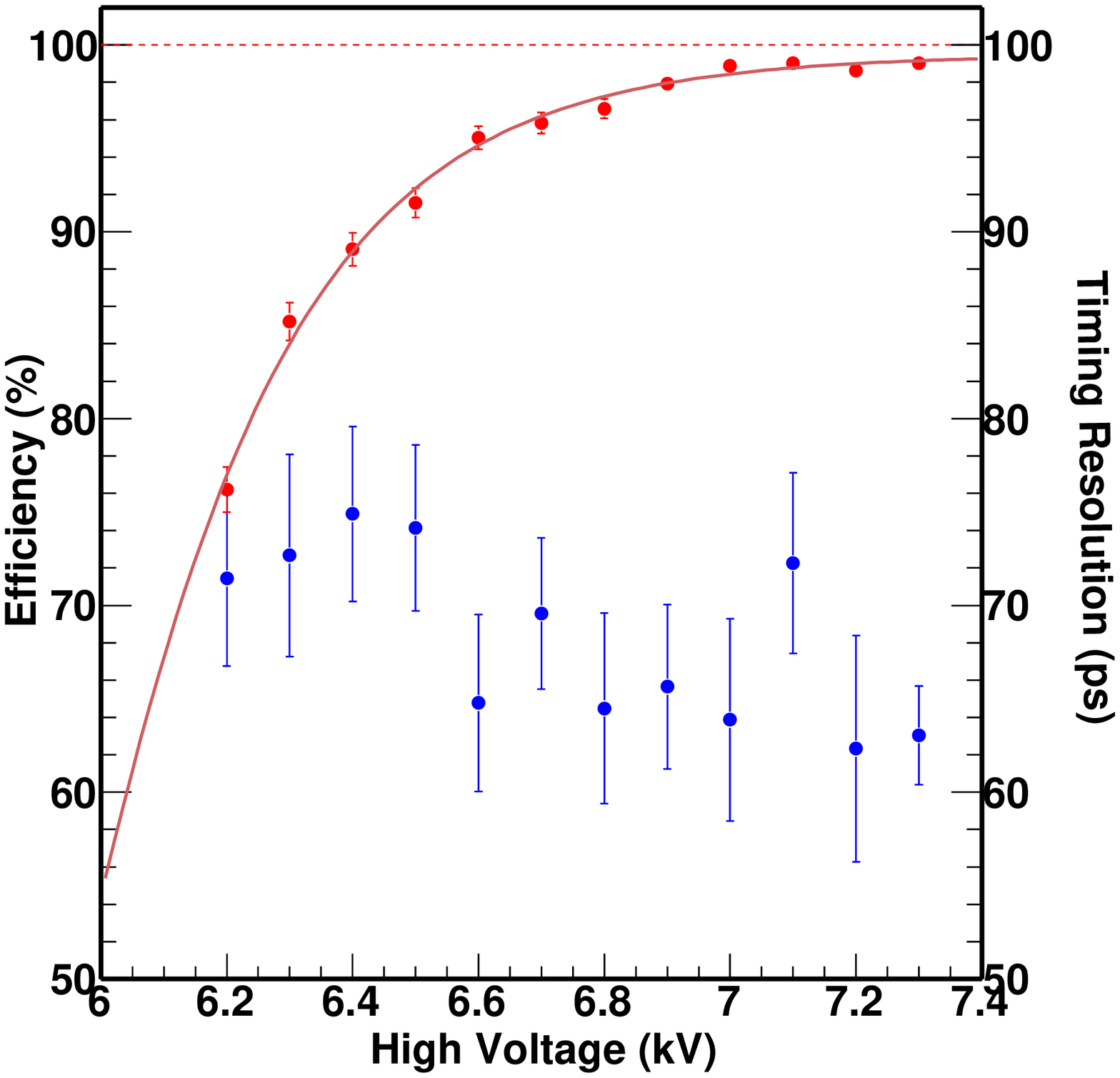}
\figcaption{\label{fig4}   The detecting efficiency and time resolution versus HV. The efficiency reaches 98\% at HV=$\pm$6.9~kV. The
time resolution, after removing the contribution from $T_0$, is between 60$\sim$70~ps in a wide range. The time resolution results include the momentum dispersion of the secondary particles. }
\end{center}

The strips are read out at each end; thus, the mean time ($T_{end1}+T_{end2}$)/2 is independent of the hit position along the strip. A typical time distribution at $\pm$7.0~kV is shown in Fig.\ref{fig5}. Since the leading edge discriminating method is used for timing, a slewing correction is needed to get rid of the dependence of timing on the signal amplitude. The variation of the time measured by the LMRPC as a function of the charge~(the corresponding QDC value) is shown in Fig.\ref{fig6}. The following function has been used to fit the plot,

\begin{eqnarray}
f(A)=p_{0}+p_{1}/\sqrt{A}+p_{2}/A+p_{3}/A^{3/2}+p_{4}/A^{2} \nonumber
\end{eqnarray}

where A stands for the charge and $p_{0}$,$p_{1}$,$p_{2}$,$p_{3}$,$p_{4}$ are the fitting parameters. The corrected time distribution is shown in Fig.\ref{fig7}. Therefore the characteristic time resolution of the LMRPC is 60~ps, in which the time jitter of the reference time has been removed. The performance uniformity in the strip direction is also important for LMRPC. Due to the limited beam time, such test was carried out only on several positions. The results show that the efficiency and time resolution are uniform in the range of ¡À6cm to the strip center.
\begin{center}
\includegraphics[width=7cm]{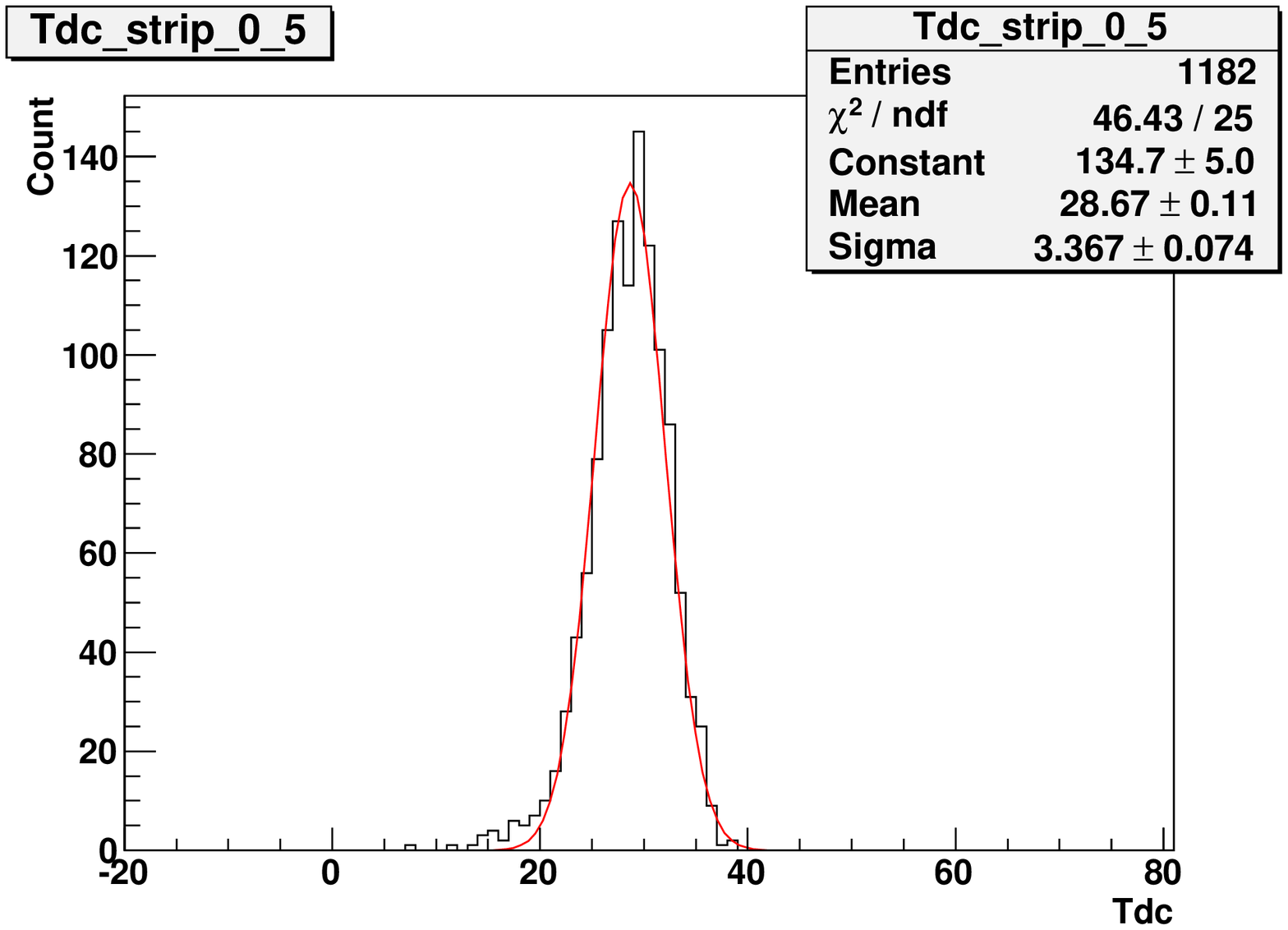}
\figcaption{\label{fig5}   The time distribution of the mean value from two ends of one strip before slewing correction. The sigma of this distribution indicates the resolution is $3.37\times 35$~ps = 118~ps~(including the $T_0$ jitter).}
\end{center}

\begin{center}
\includegraphics[width=7cm]{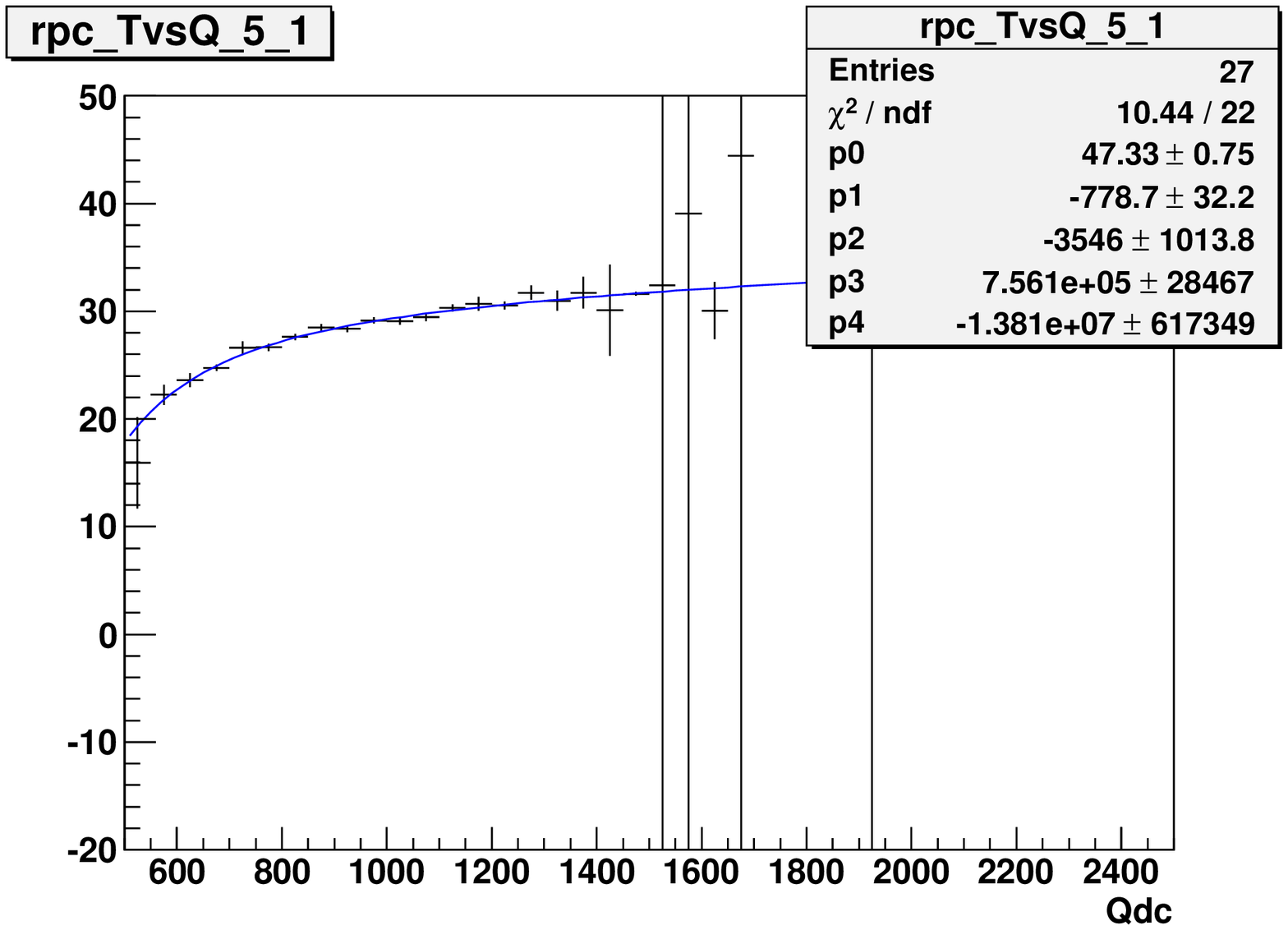}
\figcaption{\label{fig6}   The correlation between the mean time and the charge. The fit parameters are used for the slewing correction. }
\end{center}

\begin{center}
\includegraphics[width=7cm]{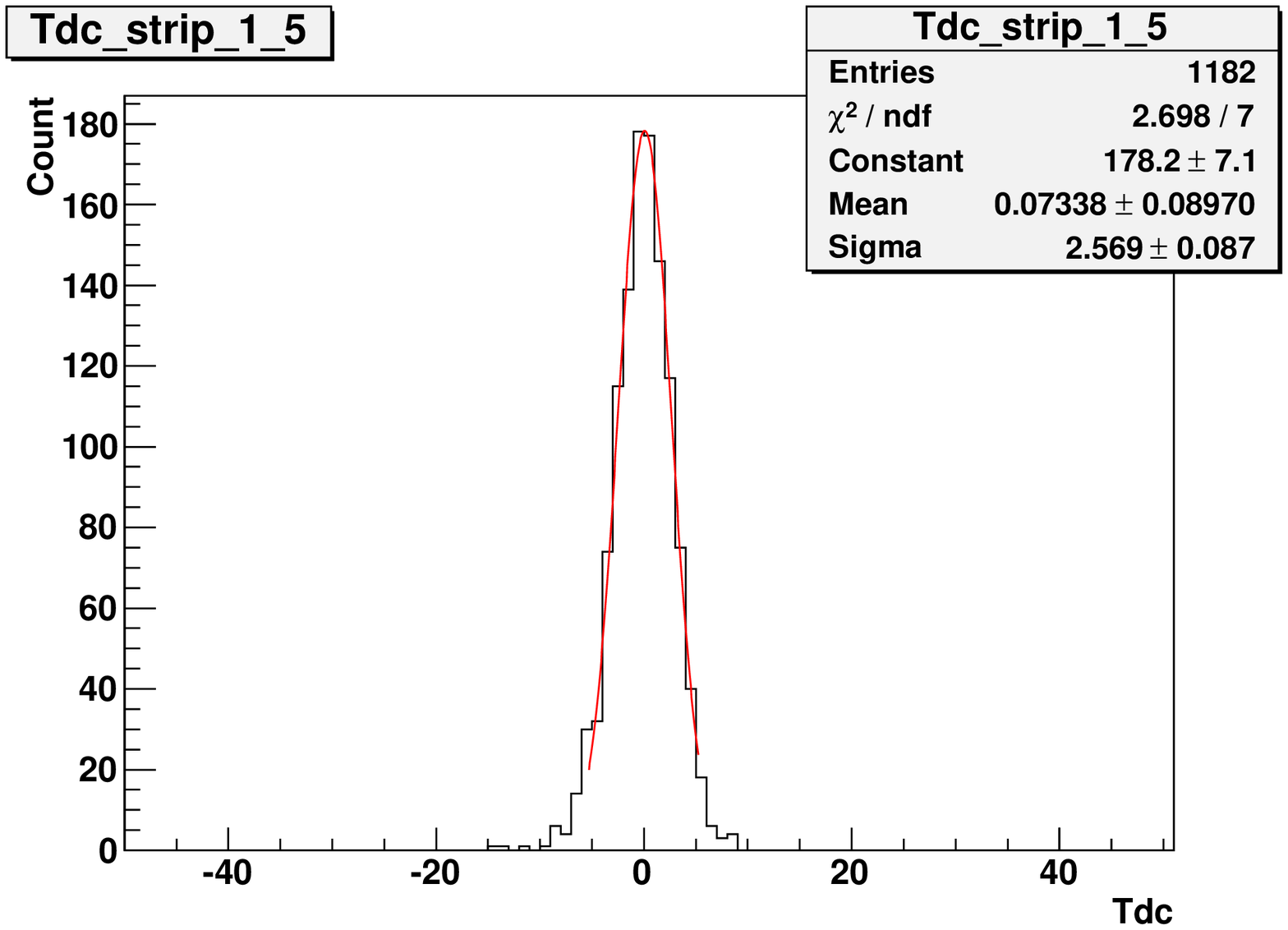}
\figcaption{\label{fig7}   The mean time distribution after slewing correction. The sigma of this distribution is 2.57 TDC channels. Removing the contribution of $T_0$~(1.91 TDC channels), the resolution of LMRPC is $\sqrt{2.57^{2}-1.91^{2}}\times35$~ps=60~ps.}
\end{center}

The LMRPC's position could be changed vertically~(along y-direction). Fig.\ref{fig8} gives the result of the detecting efficiency of all the six strips~(labeled from 1 to 6) as the function of position. The positions in the plot are the relative y-coordinates of the LMRPC to the platform, instead of the hit positions of the beam particles. Since the trigger area is 2~cm wide, which is only 0.5~cm narrower than the width of the strip, the efficiencies of adjacent strips change slowly near the strip edge. When the trigger area is in the center of one strip, few particles would hit on the adjacent strips. In this case,
the efficiencies of adjacent strips are mainly from the contribution of cross-talk. In Fig.\ref{fig8},
when the trigger area is in the center of strip 4, both the efficiencies of strip 3 and strip 5 are about 3\%.
Thus the ratio of cross-talk should be less than 3\%.

\begin{center}
\includegraphics[width=6cm]{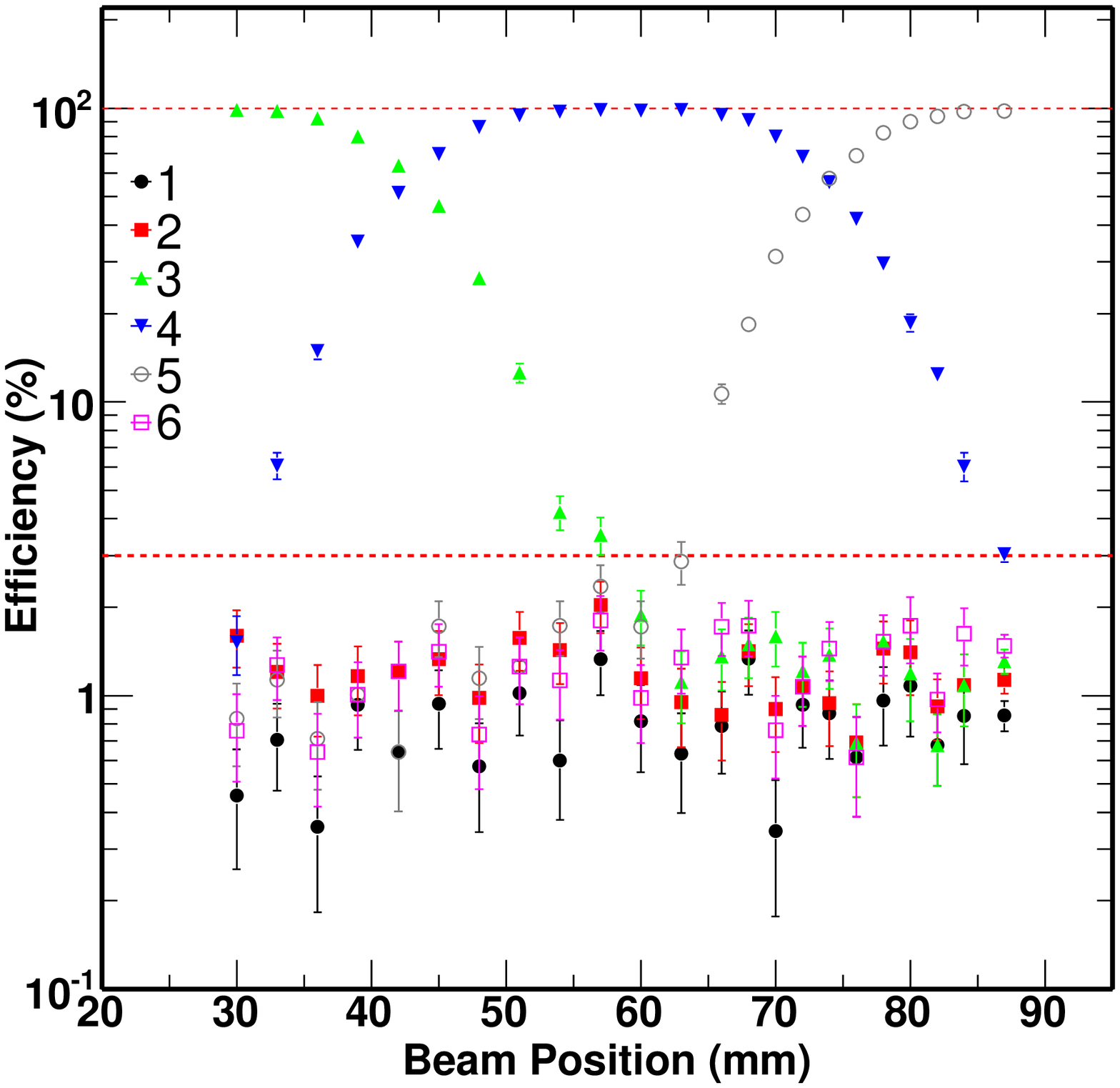}
\figcaption{\label{fig8}   The efficiency of all the six strips versus relative position. When triggered in the center of a strip,
the efficiency on adjacent strips indicates the cross-talk should be less than 3\%. }
\end{center}

\section{The position scan with the Silicon Strip Detectors~(SSD)}

To better understand the cross-talk level of the LMRPC, it's necessary to know the precise hit position of each event on the strip. At the later period of the beam test experiment, two pairs of SSDs were added to the beam test system to get the track information. The two pairs of SSDs were placed in front of the two scintillators. According to the geometry, it's very easy to reconstruct the hit position of each event on the LMRPC strip, based on the hit position recorded by the SSDs. Fig.\ref{fig9} shows the scattered distribution of reconstructed hit position on LMRPC. The trigger area is observed about 3$\times$2~$cm^{2}$.

\begin{center}
\includegraphics[width=7cm]{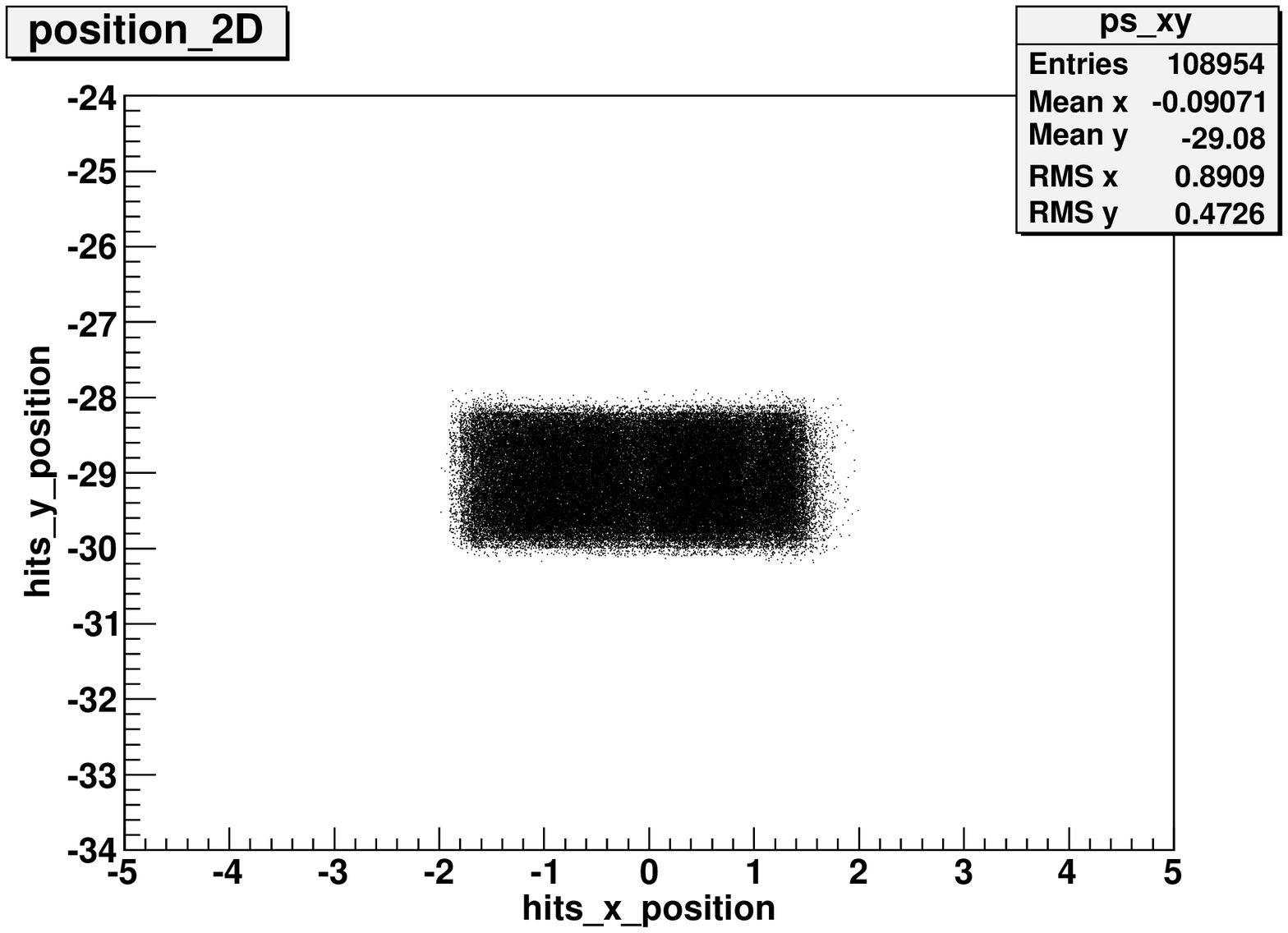}
\figcaption{\label{fig9}   The reconstructed hit position on one strip using the information given by SSDs. The trigger size on the LMRPC plane is observed to be $3\times2~cm^{2}$. }
\end{center}

\begin{figure*}
    \subfigure[]{
        \label{fig10:subfig:a}
        \begin{minipage}[b]{0.5\textwidth}
          \centering
          \includegraphics[width=6cm]{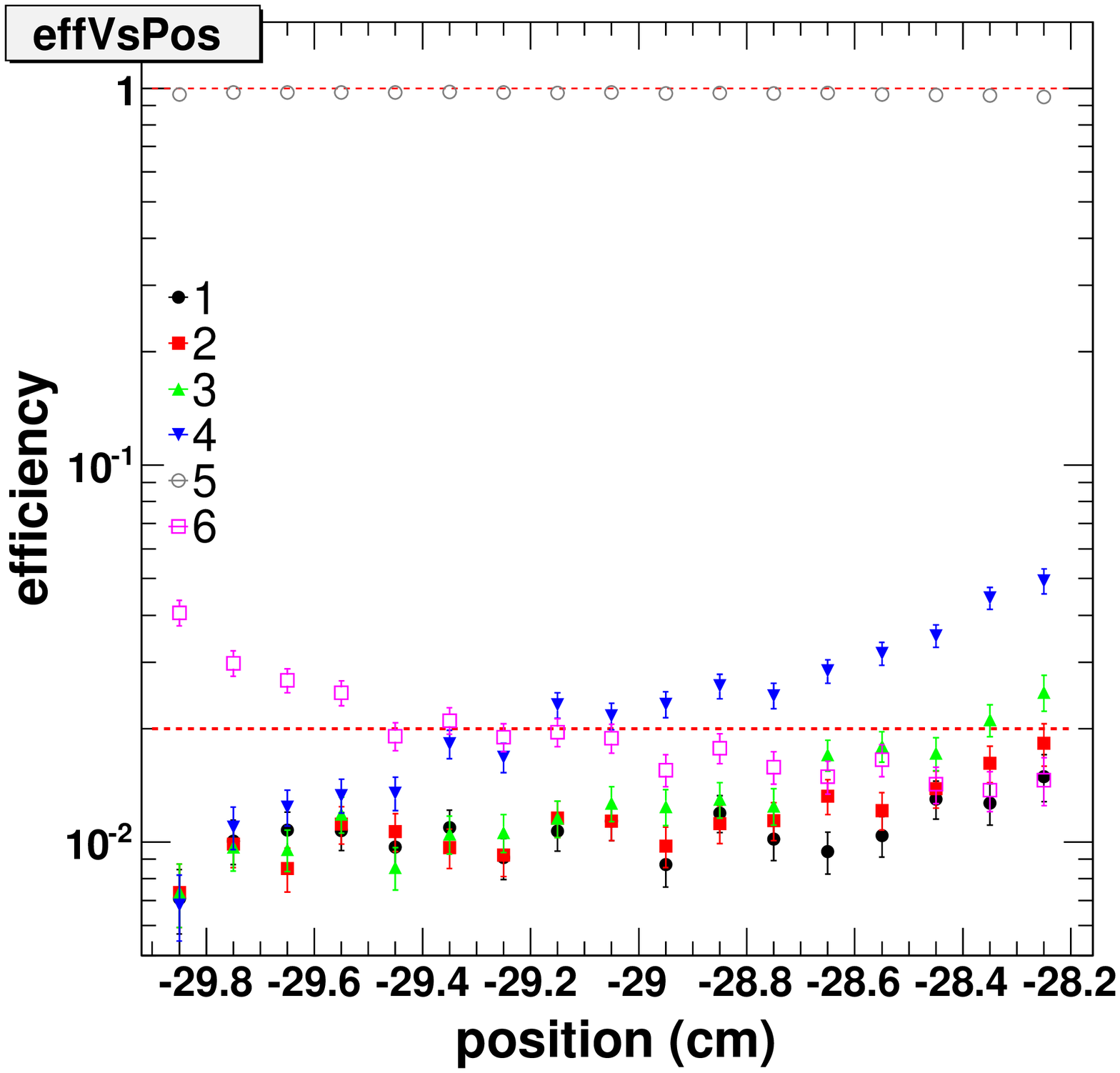}
        \end{minipage}}
    \subfigure[]{
        \label{fig10:subfig:b}
        \begin{minipage}[b]{0.5\textwidth}
          \centering
          \includegraphics[width=6cm]{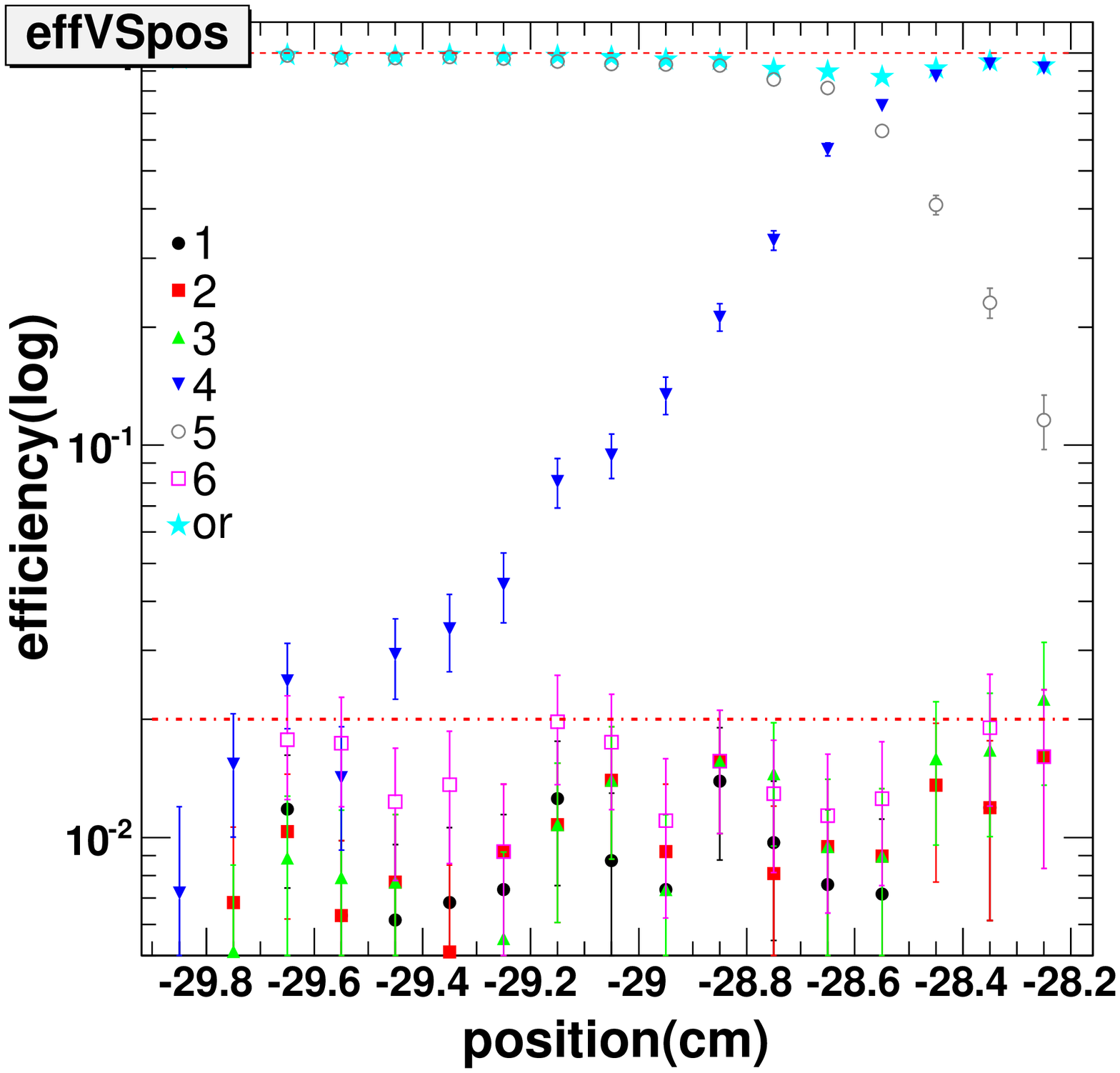}
        \end{minipage}}
    \caption{Position scan results when the trigger area was (a) in the center of one strip, (b) over the gap between two strips. The position information comes from the SSDs.}

\end{figure*}

Since the precise hit position of each event can be obtained, the ``position scan'' can be analyzed in some detail,
as shown in Fig.\ref{fig10:subfig:a} and Fig.\ref{fig10:subfig:b}. When the trigger area is in the center of one strip, say, Strip 5, which means all the beam particles hit on only one strip, the detecting efficiency of all the six strips as the function of hit y-position is shown in Fig.\ref{fig10:subfig:a}. From the edge to the center, the efficiency of Strip 4 and Strip 6 decreases slowly, which is related to the decreasing of charge sharing effect; while the efficiency of the other 3 strips remains
almost the same, since their efficiencies are mostly from cross-talk. Fig.\ref{fig10:subfig:b} is the case for the trigger area covering the gap between two strips. Approaching the edge of the strip, the LMRPC detecting efficiency does
not decrease that much, and the ``or efficiency'' is above 85\% at the center of the gap. Consistent with the
result in Fig.\ref{fig8}, the cross-talk ratio is below 2\%.

The correlation between``left minus right'' timing measured from the two ends of one LMRPC strip and the particle hit position (in x-direction, along the strip) given by SSDs is shown in Fig.\ref{fig11}. The strong correlation observed confirms that the time information can be used to calculate the position of the hit along the strip. The slope of the linear fit indicates the signal propagation velocity is$\sim$60~ps/cm. The variance around this correlation, as shown in Fig.\ref{fig12}, is assumed to result purely from the LMRPC spatial resolution and is observed to be of 0.36~cm, since the spatial resolution of the SSDs is only about 20$\mu$m.

\begin{center}
\includegraphics[width=6cm]{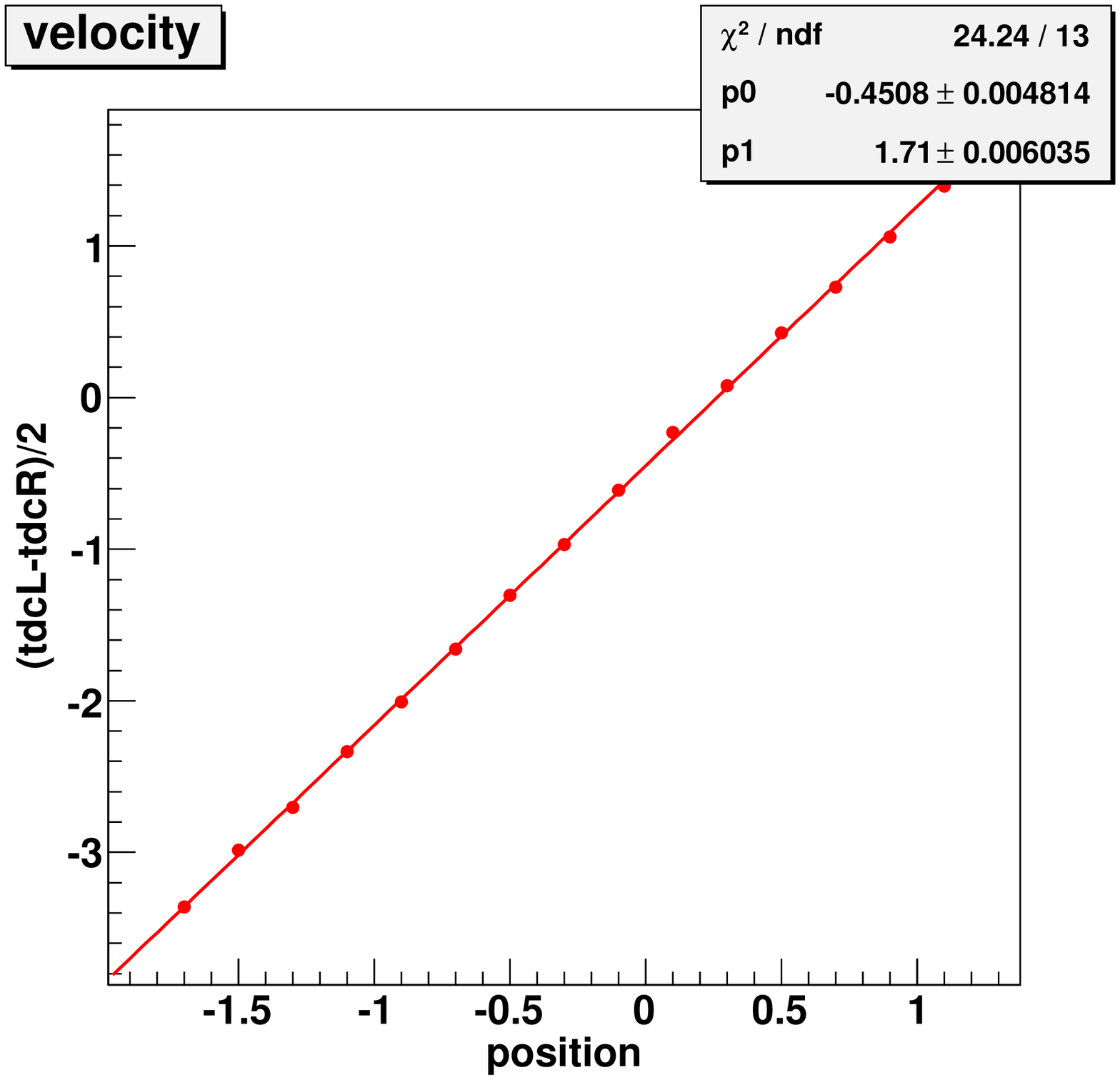}
\figcaption{\label{fig11}   The correlation between the particle position measured by SSDs and the LMRPC time difference. The linear correlation gives the signal propagation velocity along strips, which is found to be $\sim$60~ps/cm. }
\end{center}

\begin{center}
\includegraphics[width=7cm]{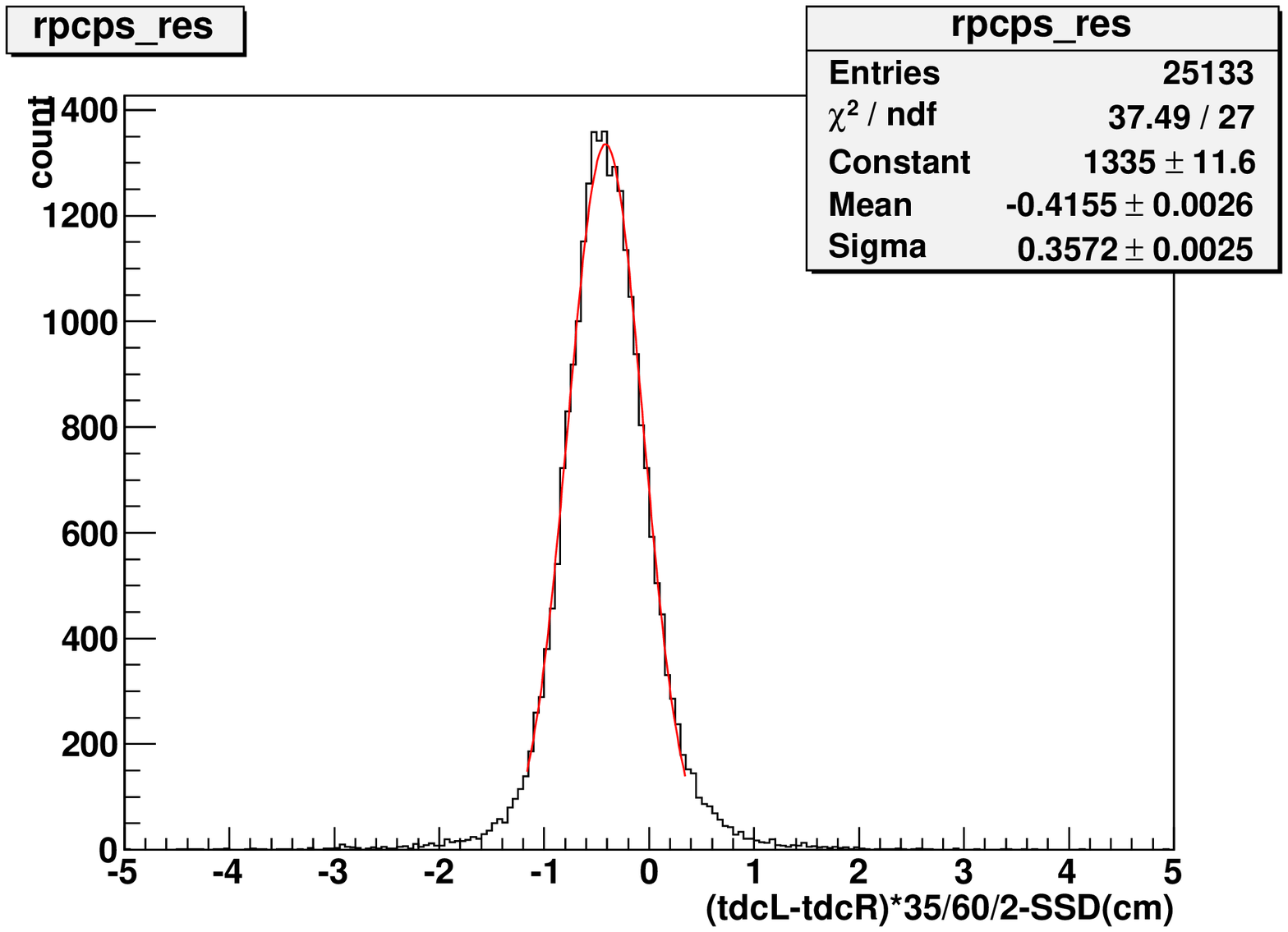}
\figcaption{\label{fig12}   With the signal propagation velocity, the hit position can be calculated by the measured time difference from two ends of strip. The deviation from the SSDs reconstructed position corresponds to the spatial resolution of the LMRPC which is $\sim$0.36~cm.  }
\end{center}

\section{Summary}

A new prototype of long-strip MRPC has been built and tested. The beam test indicates that this LMRPC prototype has a
time resolution of about 60$\sim$70~ps and a detecting efficiency greater than 98\% at a high voltage of $\pm$7~kV. The cross-talk
ratio is below 2\% and thus it indicates this prototype design can effectively reduce the cross-talk and provide a possible solution for LMRPC to work at high multiplicity. By reading out from two ends, a spatial resolution of 0.36~cm along the strip has been obtained, which can provide additional tracking information. However, due to the edge effect, the detecting efficiency at the gap between two adjacent strips decreases, which is the side effect of increasing the gap width to reduce the cross-talk. The balance between reducing the cross-talk and increasing the edge efficiency is significant to design and optimize the specialty of LMRPC.

\acknowledgments{We would like to acknowledge P. Senger, N. Herrmann, D. Gonzalez-Diaz, I. Deppner, Y. Zhang, and the CBM colleagues for their great support in the beam test. We also wish to thank the FOPI group for the support of beam operation.}

\end{multicols}

\vspace{30mm}

\begin{multicols}{2}

\end{multicols}

\clearpage

%\end{CJK*}

\end{document}